\documentclass[AER]{AEA2}%
\usepackage{harvard}
\usepackage{graphicx}%
\usepackage{amsmath}%
\usepackage{amsfonts}%
\usepackage{amssymb}
\providecommand{\U}[1]{\protect\rule{.1in}{.1in}}

\draftSpacing{1}
\begin{document}

\date{%
}
\title{Leverage Causes Fat Tails and Clustered Volatility}

\author{Stefan Thurner, J. Doyne Farmer and John Geanakoplos\thanks{ Thurner: Complex
Systems Research Group, Medical University of Vienna, W\"ahringer G\"urtel
18-20, A-1090, Austria, Santa Fe Institute, 1399 Hyde Park Road, Santa Fe, NM
87501, USA, thurner@univie.ac.at. Farmer: Santa Fe Institute, 1399 Hyde Park
Road, Santa Fe, NM 87501, USA, LUISS Guido Carli, Viale Pola 12, 00198 Roma,
Italy, jdf@santafe.edu. Geanakoplos: James Tobin Professor of Economics, Yale
University, New Haven CT, USA, Santa Fe Institute, 1399 Hyde Park Road, Santa
Fe, NM 87501, USA, john.geanakoplos@yale.edu. We thank Barclays Bank, Bill
Miller, the National Science Foundation grant 0624351 and the Austrian Science
Fund grant P17621 for support. We would also like to thank Duncan Watts for
stimulating the initiation of this project, and Alan Kirman and Chris Wood for
useful discussions. }}

\begin{abstract}

We build a simple model of leveraged asset purchases with margin calls.
Investment funds use what is perhaps the most basic financial strategy, called
\textquotedblleft value investing", i.e. systematically attempting to buy
underpriced assets. When funds do not borrow, the price fluctuations of the
asset are normally distributed and uncorrelated across time. All this changes
when the funds are allowed to \textit{leverage}, i.e. borrow from a bank, to
purchase more assets than their wealth would otherwise permit. During good
times competition drives investors to funds that use more leverage,
because they have higher profits. As leverage increases price fluctuations
become heavy tailed and display clustered volatility, similar to what is
observed in real markets. Previous explanations of fat tails and clustered
volatility depended on \textquotedblleft irrational behavior", such as trend
following. Here instead this comes from the fact that leverage limits cause
funds to sell into a falling market: A prudent bank makes itself locally safer
by putting a limit to leverage, so when a fund exceeds its leverage limit, it
must partially repay its loan by selling the asset. Unfortunately this
sometimes happens to all the funds simultaneously when the price is already
falling. The resulting nonlinear feedback amplifies large downward price
movements. At the extreme this causes crashes, but the effect is seen at every
time scale, producing a power law of price disturbances. A standard
(supposedly more sophisticated) risk control policy in which individual banks
base leverage limits on volatility causes leverage to rise during periods of
low volatility, and to contract more quickly when volatility gets high, making
these extreme fluctuations even worse.



\end{abstract}
\maketitle

\shortTitle{Leverage causes fat tails and clustered volatility}

\pubMonth{} \pubYear{} \pubVolume{} \pubIssue{}

\section{Introduction}

Recent events in financial markets have underscored the dangerous consequences
of the use of excessive credit. At the most basic level the problem is
obvious: If a firm buys assets with borrowed money, then under extreme market
conditions it may owe more money than it has and default. If this happens on a
sufficiently wide scale then it can severely stress creditors and cause them
to fail as well.

We show here that a special but extremely widespread kind of credit called
collateralized loans with margin calls has a more pervasive effect: when used
excessively it can cause default and crashes, but it also leaves a signature
even when there is no default or crash. These kinds of loans have already been
identified as a major culprit in the recent crisis, and in previous near
crises as well\footnote{For previous equilibrium-based analyses of leverage,
which show that prices crash before default actually occurs, see Geanakoplos
\cite{Geanakoplos97,Geanakoplos03,Fostel08,Geanakoplos09}. See also 
\cite{Brunnermeier09}.}. But we show here
that they create a dynamic
 in asset price fluctuations that manifests itself
at all time scales and to all degrees. The extraordinary crisis of the last
couple of years is just one extreme (but not extremal) point on a continuum.

By taking out a collateralized loan a buyer of stocks or mortgage backed
securities can put together a portfolio that is worth a multiple of the cash
he has available for their purchase. In 2006 this multiple or ``leverage"
reached 60 to 1 for AAA rated mortgage securities, and 16 to 1 for what are
now called the toxic mortgage securities. The outstanding volume of these
leveraged asset purchases reached many trillions of dollars. Leverage has
fluctuated up and down in long cycles over the last 30 years.

Conventional credit is for a fixed amount and a fixed maturity, extending over
the period the borrower needs the money. In a collateralized loan with margin
calls, the debt is guaranteed not by the reputation (or punishment) of the
borrower, but by an asset which is confiscated if the loan is not repaid.
Typically the loan maturity is very short, say a day, much shorter than the
length of time the borrower anticipates needing the money. The contract
usually specifies that after the daily interest is paid, as long as the loan
to asset value ratio remains below a specified threshold, the debt is rolled over
another day (up to some final maturity, when the threshold ratio might be
changed). If, however, the collateral asset value falls, the lender makes a
margin call and the borrower is expected to repay part of the debt and so roll
over a smaller loan to maintain the old loan to value threshold. Quite often
the borrower will obtain the cash for this extra downpayment by selling some
of the collateral. The nature of the collateralized loan contract thus
sometimes turns buyers of the collateral into sellers, even when they might
think it is the best time to buy.

Needless to say, the higher the loan to value, or equivalently, the higher the
\textit{leverage} ratio of asset value to cash downpayment, the more severe
will be the feedback mechanism. A buyer who is at his threshold of $\lambda$
times leveraged loses $\lambda\%$ of his investment for every 1\% drop in the
asset price, and on top of that will have to come up with $\$(\lambda
-1)/\lambda$ of new cash for every \$1 drop in the price of the asset. When
there is no leverage, and $\lambda=1,$ there is no feedback, but as the
leverage increases, so does the feedback.

The feedback from falling asset prices to margin calls to the transformation
of buyers into sellers back to falling asset prices creates a nonlinear
dynamic to the system. The nonlinearity rises as the leverage rises. This
nonlinear feedback would be present in the most sophisticated rational
expectations models or in the most simple minded behavioral models: it is a
mechanical effect that stems directly from the nonlinear dynamics caused by
the use of leverage and margin calls. We therefore build the simplest model
possible and then simulate it over tens of thousands of periods, measuring and
quantifying the effect of leverage on asset price fluctuations.

Our model provides a new explanation for the fat tails and clustered
volatility that are commonly observed in price fluctuations
\cite{Mandelbrot63,Engle82}. Clustered volatility and fat tails emerge in the
model on a broad range of time scales, including very rapid ones and very slow
ones. Mandelbrot and Engle found that actual price fluctuations did not
display the independent and normally distributed properties assumed by the
pioneers of classical finance \cite{Bachelier64,Black73}. Though their work
has been properly celebrated, no consensus has formed on the mechanism which
creates fat tails and clustered volatility.  The mechanism we develop here supports the
hypothesis that they are caused by the endogenous dynamics of the market rather than the nature of information itself -- in our model information is normally distributed and IID, but when leverage
is used, the resulting prices are not.

Previous endogenous explanations assume the presence of a kind of trader who
exacerbates fluctuations. \ Traders in these models are of at least two types:
value investors, who make investments based on fundamentals, and trend
followers, who make investments in the direction of recent price
movements\footnote{See
\cite{Palmer94,Arthur97,Brock97,Brock98,Lux98,Lux99,Caldarelli97,Giardina03}.
See also \cite{Friedman07}, who induce bubbles and crashes via myopic learning
dynamics.}. Trend followers are inherently destabilizing, and many would
dispute whether such behavior is rational. Value investors, in contrast, are
essential to maintain a reasonably efficient market: They gather information
about valuations, and incorporate it into prices. Thus in this sense value
investing is rational. In typical models of this type, investors move their
money back and forth between trend strategies and value strategies, depending
on who has recently been more successful, and fat tails and clustered
volatility are generated by temporary increases in destabilizing trend
strategies. The mechanism that we propose here for fat tails and clustered
volatility only involves value investors, who are stabilizing in the absence of leverage.  With leverage, however, clustered volatility and fat tails emerge on a broad
range of time scales, including very rapid ones; our explanation has the
advantage that it can operate on such time scales (whereas it is not obviously
plausible that real agents switch from value investing to trend following at
the rapid speed that is required to explain the heavy tails which are observed on timescales going down to minutes \cite{Plerou99}).

An important aspect of our model is that even though the risk control policies
used by the individual bank lenders are reasonable from a narrow, bank-centric
point of view, when a group of banks inadvertently acts together, they can
dramatically affect prices, inducing nonlinear behavior at a systemic level
that gives rise to excessive volatility and even crashes. Attempts to regulate
risk without taking into account systemic effects can backfire, accentuating
risks or even creating new ones.\footnote{Another good example from the recent
meltdown illustrating how individual risk regulation can create systemic risk
is the use of derivatives.}  

The wealth dynamics in our model also illustrates the interaction between evolutionary dynamics that occur on very long time scales, and short term dynamics that occur on timescales of minutes.  In our model different agents use different levels of leverage; everyone is initially given the same wealth.  Under these circumstances the market is stable, and agents who use more leverage produce higher returns and attract more investors.  As time goes on the most aggressive investors accumulate more wealth, and the average leverage increases.  Eventually the leverage gets so high that there is a crash. The fact that there is evolutionary pressure toward higher and higher leverage illustrates the need for regulation.

\section{The Model}

In our model, traders have a choice between owning a single asset, such as a
stock or a commodity, or owning cash. There are two types of traders, noise
traders and funds. The noise traders buy and sell nearly at random, with a
slight bias that makes the price weakly mean-revert around a perceived
fundamental value $V$. The funds use a strategy that exploits mispricings by
taking a long position (holding a net positive quantity of the asset) when the
price is below $V$, and otherwise staying out of the market. The funds can
augment the size of their long position by borrowing from a bank at an
interest rate that for simplicity we fix at zero, using the asset as
collateral. This borrowing is called leverage. The bank will of course be
careful to limit its lending so that the value of what is owed is less than
the current price of the assets held as collateral. Default occurs if the
asset price falls sufficiently far before the loan comes due in the next period.

In addition to the two types of traders there is a representative investor who
either invests in a fund or holds cash. The amount she invests in a given fund
depends on its recent historical performance relative to a benchmark return
$r^{\mathrm{b}}$. Thus successful funds attract additional capital above and
beyond what they gain in the market and similarly unsuccessful funds lose
additional capital.

\subsection{Supply and Demand}

The total supply of the asset is $N.$ At the beginning of each period $t\geq1$
all agents observe the unit asset price $p(t).$ As is
traditional, all the traders in our model are perfectly competitive; they take
the price as given, imagining that they are so small that they cannot affect
the price, no matter how much they demand.

The \textit{noise traders'} demand is defined in terms of the cash value
$\xi_{\mathrm{nt}}(t)$ they spend on the asset, which follows the equation
\[
\log\xi_{\mathrm{nt}}(t)=\rho\log\xi_{\mathrm{nt}}(t-1)+\sigma\chi
(t)+(1-\rho)\log(VN),
\]
where $\chi$ is normally distributed with mean zero and standard deviation
one. The noise traders' demand is
\[
D_{\mathrm{nt}}(t,p(t))=\frac{\xi_{\mathrm{nt}}(t)}{p(t)} .
\]
If there were no other participants in the market, the price would be set such that
$D_{\mathrm{nt}}(t,p(t))=N$. This choice of the noise trader process would
then guarantee, with $\rho<1,$ that the price would be a mean reverting random
walk with $E[\log p]=\log V$. In the limit as $\rho\rightarrow1$ the log
returns $r(t)=\log p(t+1)-\log p(t)$ would be normally distributed. For the
purposes of this paper we fix $V=1$, $N=1000$, $\sigma=0.035$ and $\rho=0.99$. Thus in
the absence of the funds the log returns are close to being normally
distributed, with tails that are slightly truncated (and hence even thinner
than normal) due to the mean reversion.

\begin{figure}
\begin{center}
\includegraphics[width=7.0cm]{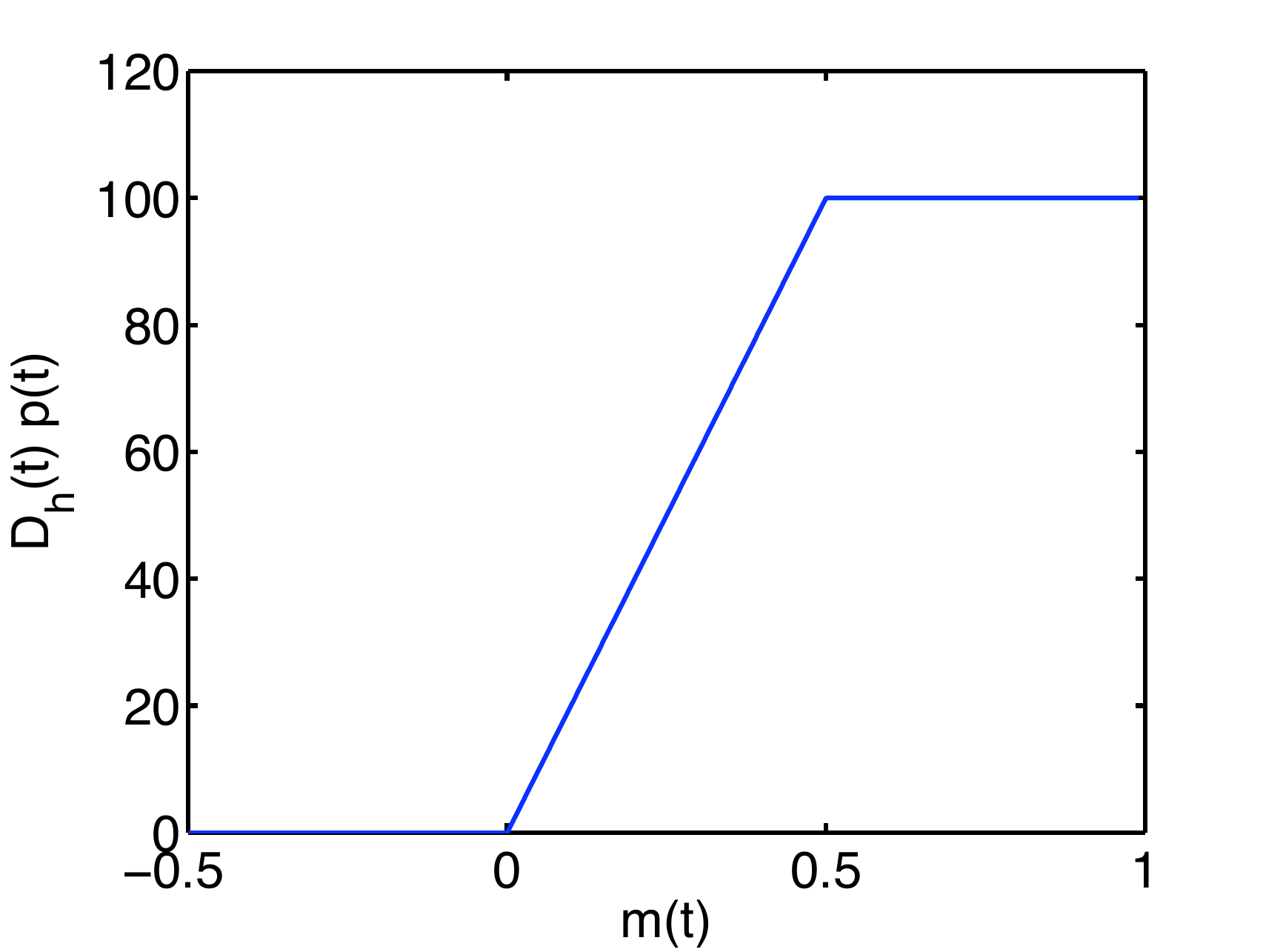}
\end{center}
\caption{Demand function $D_{h}(t)p(t)$ of a fund (measured in dollars) vs. the
mispricing signal $m(t) = V - p(t)$.  
The investor does nothing when the asset is overpriced, and she buys more and 
more as the asset becomes underpriced, until she hits her leverage limit at $m = m^{crit}$.  
After this her demand remains flat.
\label{valueInvestor}
}
\end{figure}

We add a second class of demanders called funds. The funds in our
model are \emph{value investors} who base their demand $D_{h}(t)$ on a
mispricing signal $m(t)=V-p(t)$. The perceived fundamental value $V$ is held
constant and is the same for the noise traders and for all funds. As shown in
Figure~\ref{valueInvestor}, each fund $h$ computes its demand $D_{h}(t)$ based
on the mispricing. As the mispricing increases, the dollar value
$D_{h}(t)p(t)$ of the asset the fund wishes to hold increases linearly, but
the position size is capped when the fund reaches the maximum leverage. Funds
differ only according to an aggression parameter $\beta_{h}$ that represents
how sensitive their response is to the signal $m.$

Each fund begins with the same wealth ${W}_{h}(0)=2$. After noting the price
$p(t),$ at each date $t\geq1$ hedge fund $h$ computes its wealth
$W_{h}(t)=W_{h}(t,p(t))$, as described in the next section. The fund must
split its wealth $W_{h}(t)$ between cash $C_{h}(t)$ and the value of the asset
$D_{h}(t)p(t)$
\[
W_{h}(t)=W_{h}(t,p(t))=D_{h}(t)p(t)+C_{h}(t)
\]
When the fund is borrowing money, $C_{h}(t)$ is negative and represents the
loan amount. If $W_{h}(t,p(t))\geq0$, the fund's demand $D_{h}(t)=D_{h}(t,p(t))$ can be written:
\begin{align}
\label{fundDemand1}
m(t)<0  & :D_{h}(t)=0\\
0<m<m_{h}^{\mathrm{crit}}  & :D_{h}(t)=\beta_{h}m(t)W_{h}(t)/p(t)\\
\label{fundDemand3}
m\geq m_{h}^{\mathrm{crit}}  & :D_{h}(t)=\lambda_{\mathrm{MAX}}W_{h}(t)/p(t).
\end{align}
In (1) the asset is over-priced and the fund holds nothing. In (2) the asset
is underpriced but the mispricing is not too large. The fund takes a position
whose monetary value is proportional to the mispricing $m(t)$, the fund's
wealth $W_{h}(t)$, and the aggression parameter $\beta_{h}$, which can vary
from fund to fund. In (3) the asset is even more underpriced so that the fund
has reached its maximum leverage $\lambda_{h}(t)=\lambda_{\mathrm{MAX}}$. This
occurs when $m(t)\geq m_{h}^{\mathrm{crit}}=\lambda_{\mathrm{MAX}}/\beta_{h}$.

The \textit{leverage} $\lambda_{h}$ is the ratio of the dollar value of the
fund's asset holdings to its wealth, i.e. 
\begin{equation}
\lambda_{h}(t)= \frac{D_{h}(t)p(t)}{W_{h}(t)}= \frac{D_{h}(t)p(t)}{(D_{h}(t)p(t)+C_{h}(t))}.
\label{leverageEQ} 
\end{equation}
The firm is required by the bank it borrows from to maintain $\lambda_{h}(t)\leq
\lambda_{\mathrm{MAX}}.$ If $\bar{\lambda}_{h}(t)=D_{h}(t-1)p(t)/W_{h}%
(t)>\lambda_{\mathrm{MAX}},$ the firm will have to sell the asset in order to
bring leverage $\lambda_{h}(t)$ under the maximum allowed. This is called
meeting a margin call.

Note that a $k\%$ change in the asset price from $p(t-1)$ to $p(t)$ causes a
$\bar{\lambda}_{h}(t)k\%$ change in wealth $W_{h}(t)$, hence the name
\textquotedblleft leverage". A firm that satisfied its leverage limit at time
$t-1$ might face a margin call at time $t$ either because $\bar{\lambda}%
_{h}(t)>1$ and $p(t)$ falls below $p(t-1),$ causing $W_{h}(t)$ to fall by a
larger percentage than the asset price, or because $W_{h}(t)$ falls below
$W_{h}(t-1)$ due to redemptions, described in the next section.

If $W_{h}(t)<0,$ the fund defaults and goes out of business. The fund sells
all its assets, demanding $D_{h}(t)=0,$ and returns all the revenue to pay off
as much of its borrowed money as it can to its bank lender. The bank bears the
loss of the default. For simplicity, we assume the bank has deep pockets and,
despite the loss, continues to lend to other funds as before. After a period
of time has passed, the defaulting fund reemerges again as a new fund, as we
shall describe below.

Prices are set by equating the demand of the funds plus the noise traders to
the fixed supply of the asset
\[
D_{\mathrm{nt}}(t,p(t))+\sum_{h}D_{h}(t,p(t))=N .
\]

\subsection{Fund Wealth Dynamics}

The funds' wealth automatically grows or shrinks according to the success or
failure of their trading. In addition it changes due to additions or
withdrawals of money by investors, as described below.  If a
fund's wealth goes below a critical threshold, here set to $W_{h}(0)/10
$, the fund goes out of business\footnote{Using a positive survival threshold
for removing funds avoids the creation of \textquotedblleft zombie funds" that
persist for long periods of time with almost no wealth.}, and after a period
of time, $T_{\rm reintro}$, has passed it is replaced by a new fund with wealth $W_{h}(0)$
and the same parameters $\beta_{h}$ and $\lambda_{\mathrm{MAX}}$. In the simulations we use 
$T_{\rm reintro}=100$ timesteps. 

A pool of fund investors, who are treated as a single representative investor, contribute or
withdraw money from each fund based on a moving average of its recent
performance. This kind of behavior is well documented\footnote{Some of the
references that document or discuss the flow of investors in and out of mutual
funds include [\cite{Busse01,Chevalier97,Delguercio02,Remolona97,Sirri98}].}.
We introduce it into our model because it guarantees a steady-state behavior
with well-defined long term statistical averages of the wealth of the hedge funds, i.e.
it allows the wealth to fluctuate but prevents it from growing without bound.

Let
\[
r_{h}(t) = \frac{D_{h}(t-1)(p(t)-p(t-1))}{W_{h}(t-1)}
\]
be the rate of return by fund $h$ on investments at time $t$. The investors make
their decisions about whether to invest in the fund based on $r_{h}%
^{\mathrm{perf}}(t)$, an exponential moving average of these performances,
defined as
\begin{equation}
r_{h}^{\mathrm{perf}}(t)=(1-a)\,r_{h}^{\mathrm{perf}}(t-1)+a\,r_{h}(t).
\end{equation}
The flow of capital in or out of the fund, $F_{h}(t)$, is given by%
\begin{align}
\tilde{F}_{h}(t) &  =b\,\,[r_{h}^{\mathrm{perf}}(t)-r^{\mathrm{b}
}]\,\,[D(t-1)p(t)+C(t-1)]\\
F_{h}(t) &  =\max(\tilde{F}_{h}(t),-[D(t-1)p(t)+C(t-1)]),
\end{align}
where $b$ is a parameter controlling the fraction of capital withdrawn and
$r^{\mathrm{b}}$ is the benchmark return of the investors. The investors
cannot take out more money than the fund has. 
\begin{figure}
\begin{center}%
\begin{tabular}
[c]{c}%
\includegraphics[width=13.0cm]{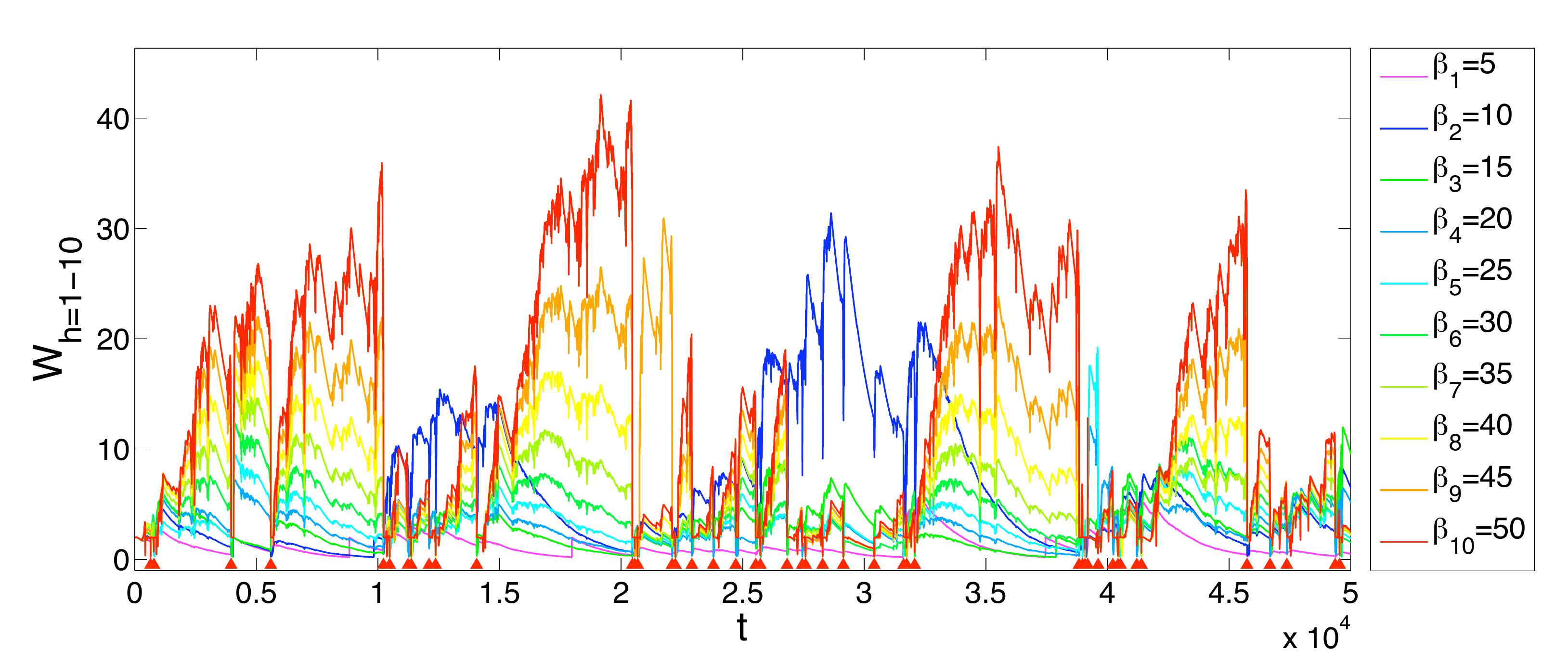}
\end{tabular}
\end{center}
\caption{Wealth timeseries $W_{h}(t)$ for 10 funds with $\beta_{h} = 5, 10, \ldots,
50$ and $\lambda_{\mathrm{MAX}}=20$ for all funds. Times at which (at
least) one fund collapses are marked by triangles. \label{eco}}
\end{figure}

Funds are initially given wealth $W_{0}=W_{h}(0)$. At the beginning of each
new timestep $t\geq1$, the wealth of the fund changes according to
\begin{equation}
W_{h}(t)=W_{h}(t-1)+[p(t)-p(t-1)]D_{h}(t-1)+F_{h}(t).
\end{equation}
In the simulations in this paper, unless otherwise stated we set $a=0.1$,
$b=0.15$, $r^{\mathrm{b}}=0.005$, and $W_{0}=2$.

The benchmark return $r^{\mathrm{b}}$ plays the important role of determining the relative size of hedge funds vs. noise traders.  If the
benchmark return is set very low then funds will become very wealthy and will
buy a large quantity of the asset under even small mispricings, preventing the
mispricing from ever growing large. This effectively induces a hard floor on
prices. If the benchmark return is set very high, funds accumulate little
wealth and play a negligible role in price formation. The interesting behavior
is observed at intermediate values of $r^{\mathrm{b}}$ where the funds' demand
is comparable to that of the noise traders.
\begin{figure}
\begin{center}
\includegraphics[width=6.0cm]{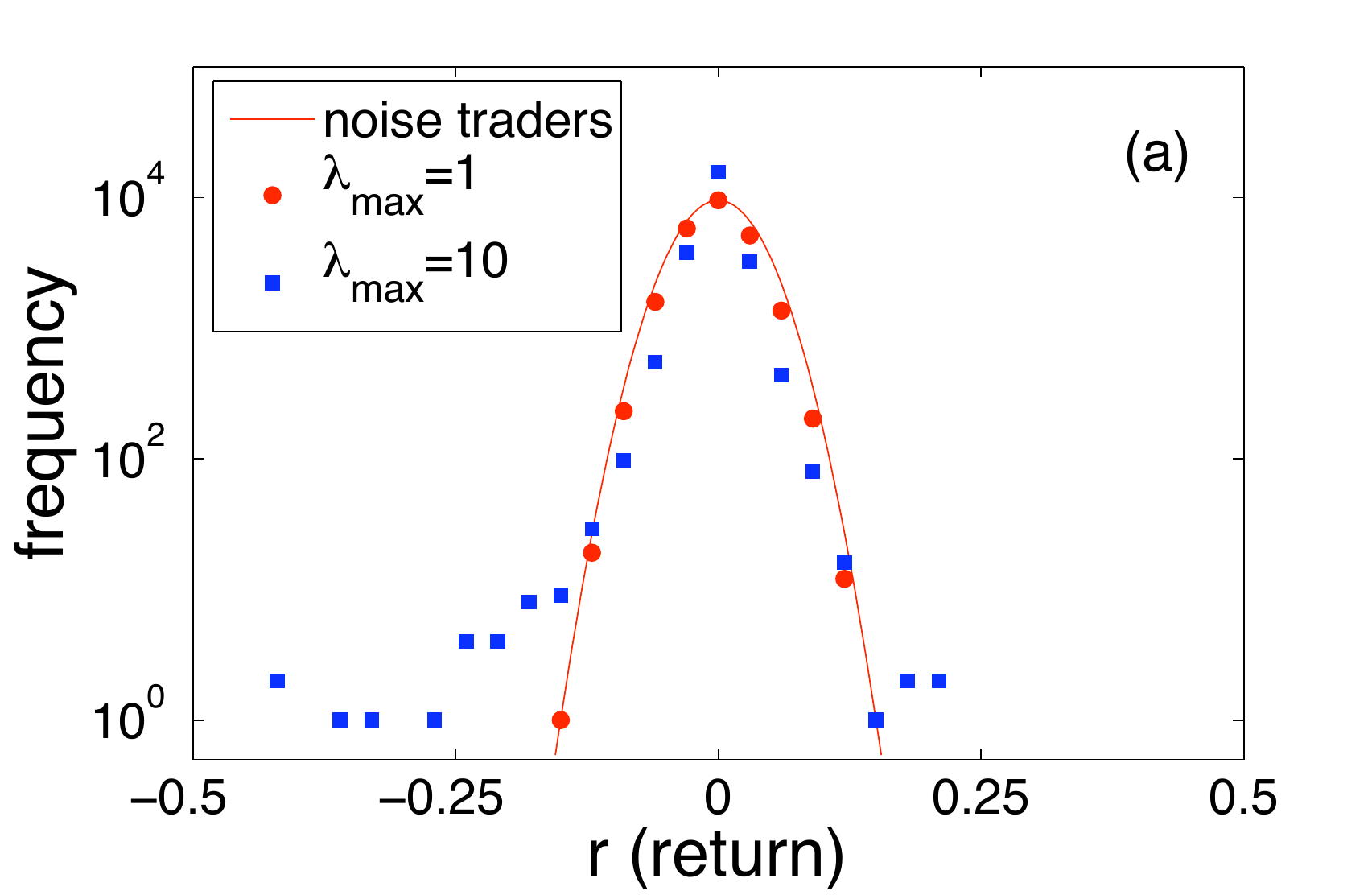}
\includegraphics[width=6.0cm]{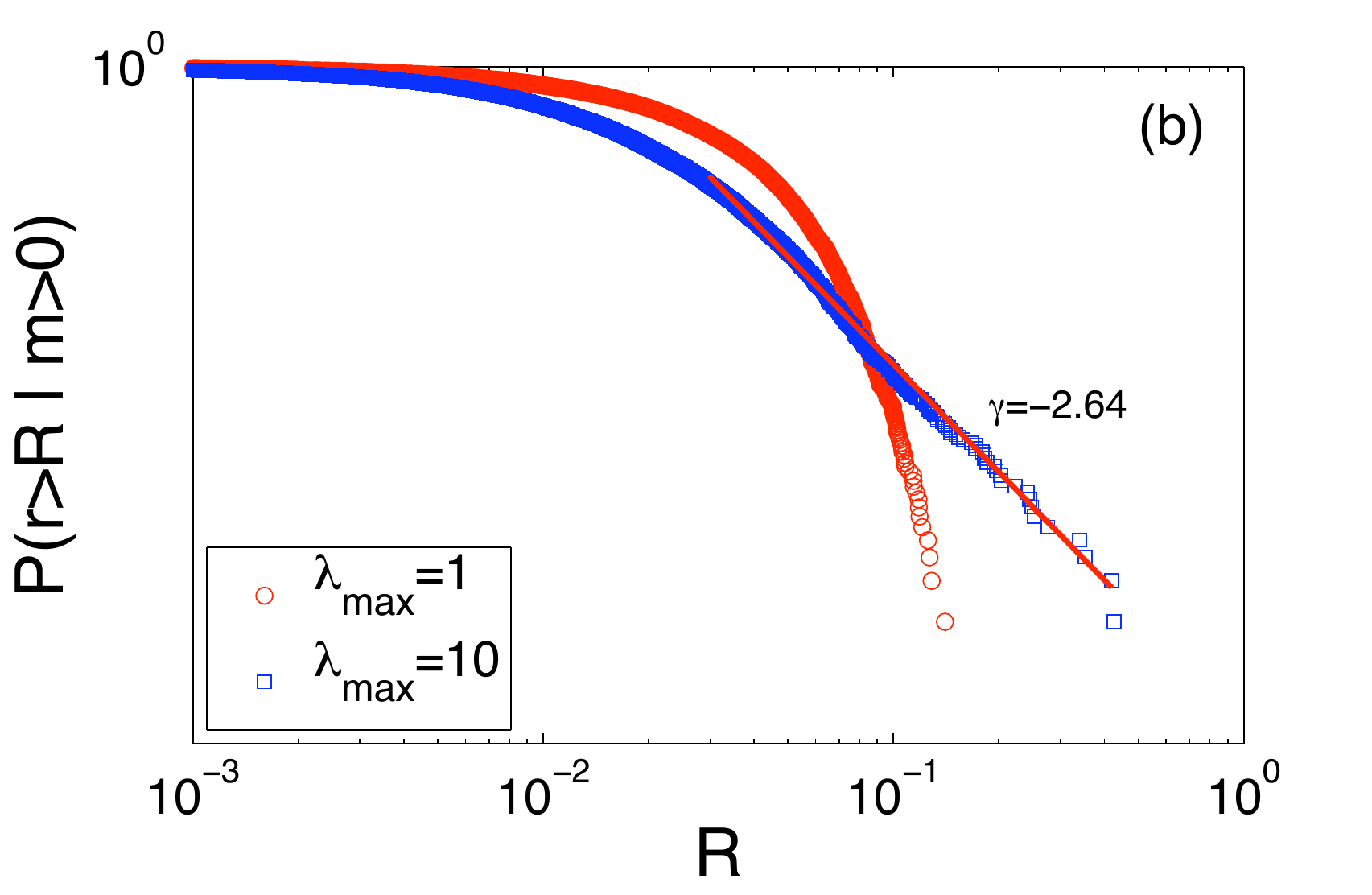}
\includegraphics[width=6.0cm]{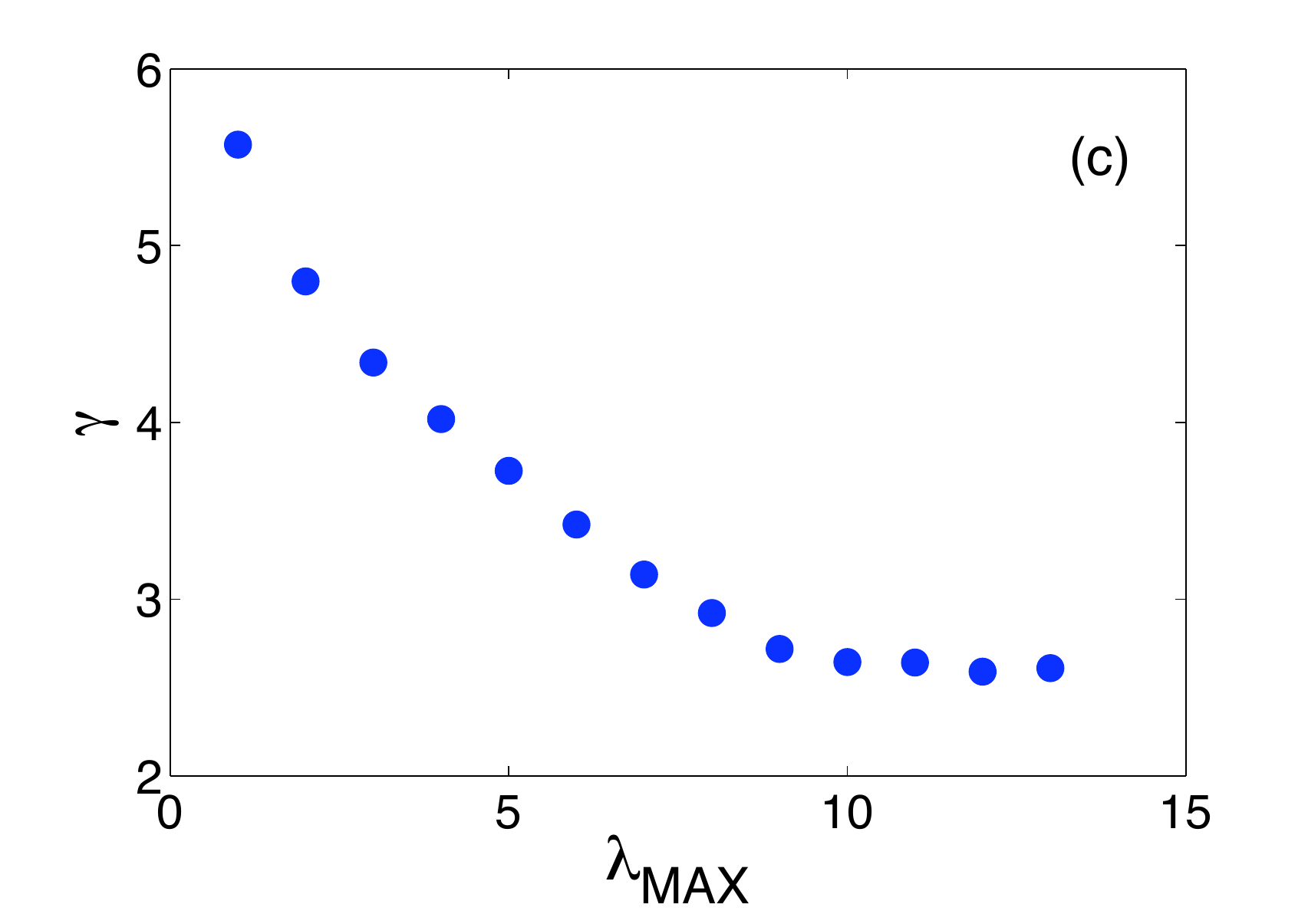}
\end{center}
\caption{The distribution of log returns $r$. (a) plots the density of log returns
$p(r | m > 0)$ on semi-log scale. The results are conditioned on positive
mispricing $m > 0$, i.e. only when the funds are active. The unleveraged case (red
circles) closely matches the noise trader only case (red curve). When the
maximum leverage is raised to ten (blue squares) the body of the distribution
becomes thinner but the tails become heavy on the negative side. This is seen
from a different point of view in (b), which plots the cumulative distribution
for negative returns, $P(r > R | m > 0)$, in log-log scale. For $\lambda_{\mathrm{MAX}}=10$ 
we fit a power law to the data across the indicated region
and show a line for comparison. In (c) we vary $\lambda_{\mathrm{MAX}}$ and
plot fitted values of $\gamma$, illustrating how the tails become heavier as
the leverage increases.  Same $\beta$ values as in Figure \ref{eco}. \label{returnDist}}
\end{figure}

\section{Simulation results}

\subsection{Wealth dynamics}

In Fig.~\ref{eco} we illustrate the wealth dynamics for a simulation with $10
$ funds whose aggression parameters are $\beta_{h}=5,10,\ldots,50$. They all
begin with the same low wealth $W_h(0)=2$; at the outset they make good returns
and their wealth grows quickly. This is particularly true for the most
aggressive funds; with higher leverage they make higher returns so long as the
asset price is increasing. As their wealth grows the funds have more impact,
i.e. they themselves affect prices, driving them up when they are buying and
down when they are selling. This limits their profit-making opportunities and
imposes a ceiling of wealth at about $W=40$. There are a series of crashes
which cause defaults, particularly for the most highly leveraged funds. Twice
during the simulation, at around $t=10,000$ and $25,000$, crashes wipe out all
but the two least aggressive funds with $\beta_{h}=5,10$. While funds
$\beta_{3}-\beta_{10}$ wait to get reintroduced, fund $\beta_{2}$ manages to
become dominant for extended periods of time.

\begin{figure}
\begin{center}
\includegraphics[width=5.4cm]{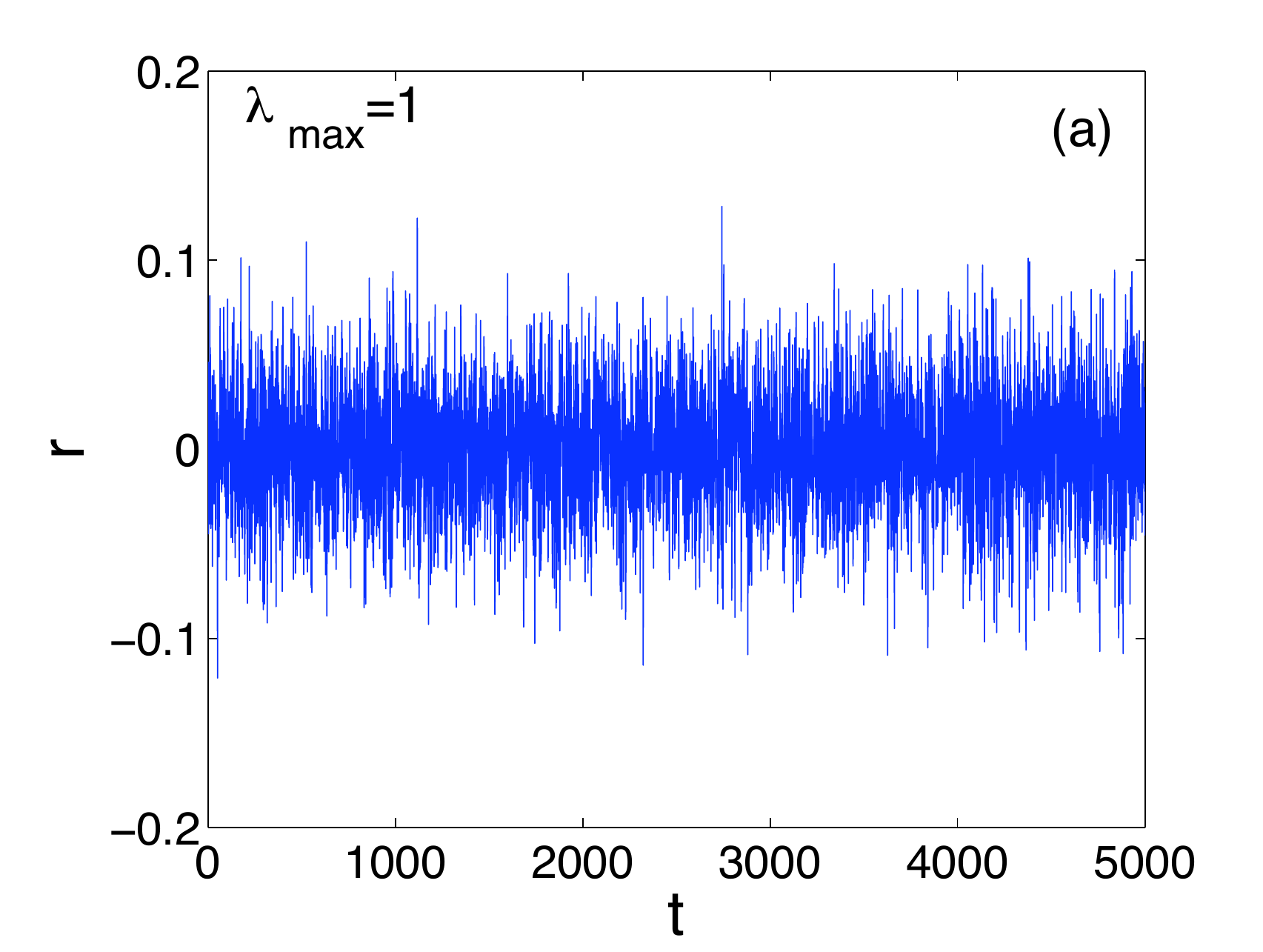}
\includegraphics[width=5.4cm]{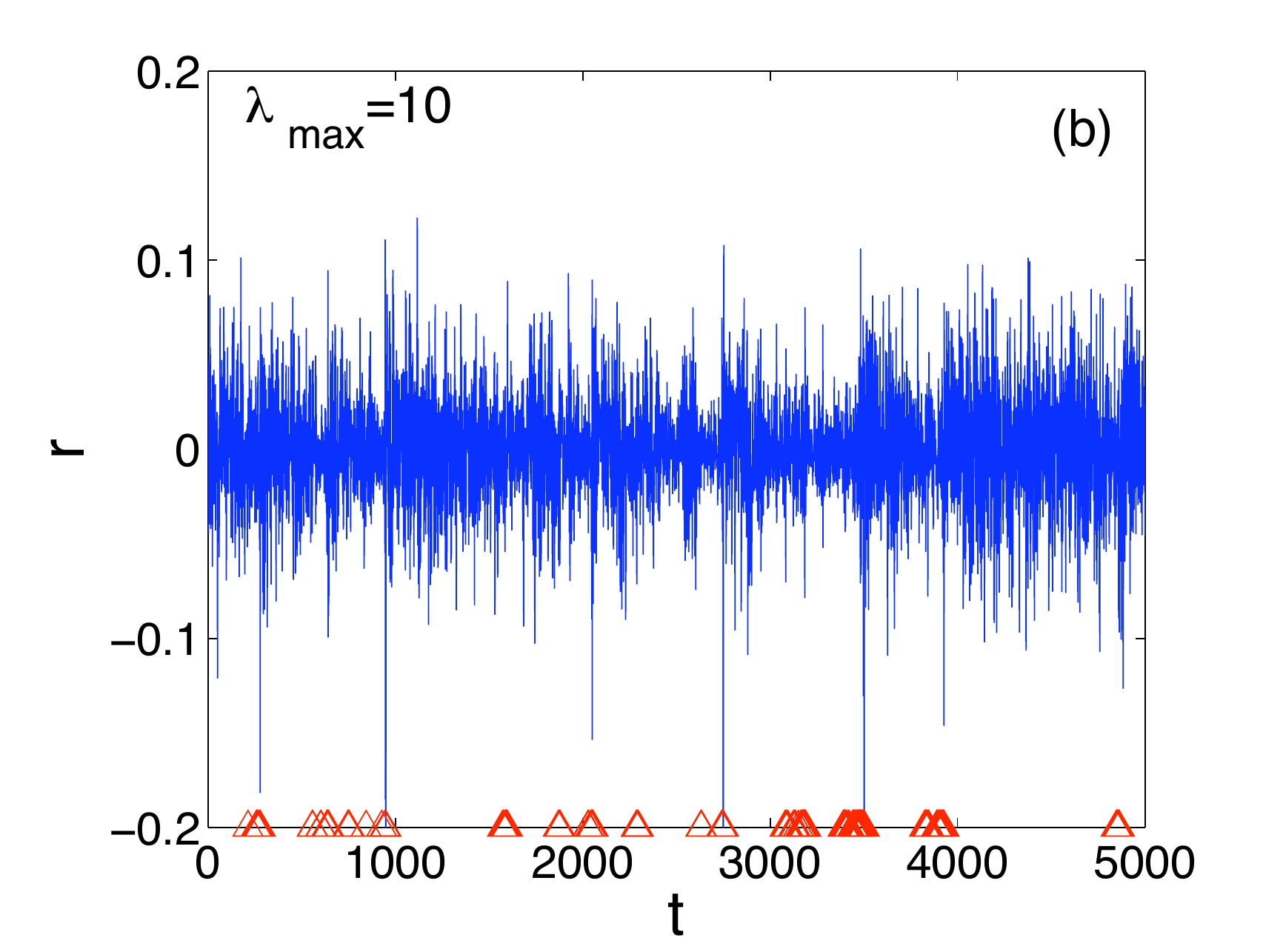}
\includegraphics[width=5.3cm]{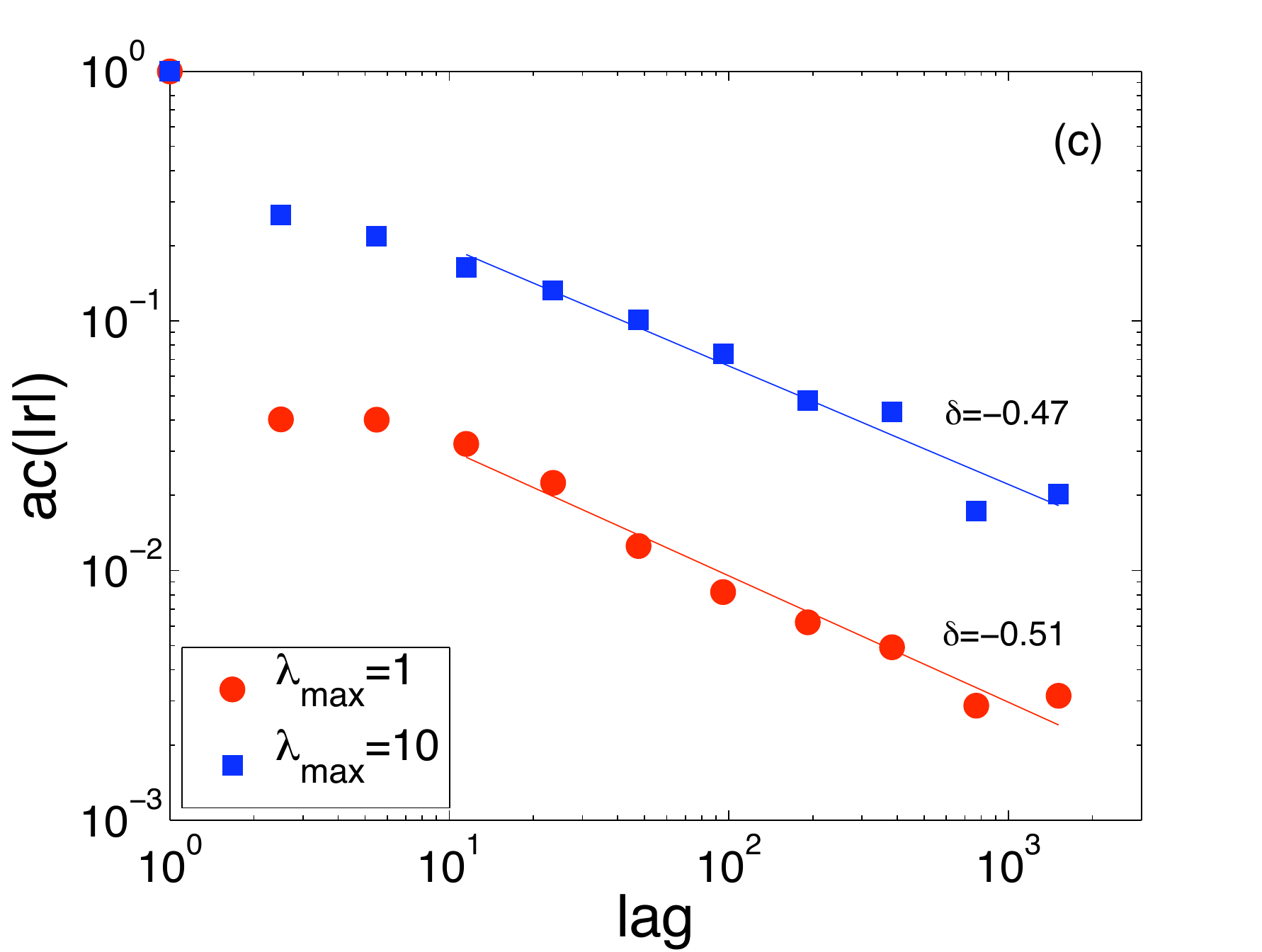}
\end{center}
\caption{ Log-return timeseries (a) $\lambda_{\mathrm{MAX}} = 1$; (b) $\lambda_{\mathrm{MAX}}=10$. 
Triangles mark margin calls in the simulation, indicating
a direct connection between large price moves and margin calls. (c)
Autocorrelation function of the absolute values of log-returns for (a-b)
obtained from a single run with $100,000$ timesteps. This is plotted on
log-log scale in order to illustrate the power law tails. (The autocorrelation
function is computed only when the mispricing is positive.) Same $\beta$
values as in Figure \ref{eco}. \label{price_ts}}
\end{figure}

\subsection{Returns and correlations}

The presence of the funds dramatically alters the statistical properties of
price returns. This is illustrated in Fig.~\ref{returnDist}, where we compare
the distribution of logarithmic price returns $r(t) = \log p(t) - \log
p(t-1)$, for three cases: (1) Noise traders only. (2) Hedge funds with no
leverage ($\lambda_{\mathrm{MAX}} = 1$). (3) Substantial leverage, i.e.
$\lambda_{\mathrm{MAX}} = 10$. With only noise traders the log returns are (by
construction) nearly normally distributed. When funds are added without
leverage the volatility of prices drops slightly, but the log returns remain
approximately normally distributed. When we increase leverage to
$\lambda_{\mathrm{MAX}} = 10$, however, the distribution becomes much more
concentrated in the center and the negative returns develop fat tails. (Recall
that since the funds are long-only, they are only active when the asset is
undervalued, i.e. when the mispricing $m > 0$. This creates an asymmetry
between positive and negative returns.) As shown in Fig.~\ref{returnDist}(b),
for $\lambda_{\mathrm{MAX}} = 10$ the cumulative distribution for the largest
negative returns roughly follows a straight line in a double logarithmic
scale, suggesting that it is reasonable to approximate the tails of the
distribution as a power law, of the form $P(r > R | m > 0) \sim R^{-\gamma}$.

The exponent $\gamma$ may be regarded as a measure of the concentration of
extreme risks. The transition from normality to fat tails occurs more or less
continuously as $\lambda_{\mathrm{MAX}}$ varies. This is in contrast to the
conjecture of Plerou et al.\cite{Plerou99,Gabaix03,Gabaix06,Plerou08} that
$\gamma$ has a universal value $\gamma\approx3$. In Figure~\ref{returnDist}(c)
we measure $\gamma$ as a function of $\lambda_{\mathrm{MAX}}$. As
$\lambda_{\mathrm{MAX}}$ increases $\gamma$ decreases, corresponding to
heavier tails\footnote{The value of $\gamma$ when $\lambda= 1$ should be
infinite, in contrast to the measured value. Large values of $\gamma$ are
difficult to measure correctly, whereas small values are measured much more
accurately.}. This trend continues until $\lambda_{\mathrm{MAX}}\approx10$,
where $\gamma$ reaches a floor at $\gamma\approx2.5$.  (The reason this floor exists depends on the particular choice of parameters here, and will be explained later).  A typical value measured
for financial time series, such as American stocks \cite{Plerou99,Cont00}, is
$\gamma\approx3$. In our model this corresponds to a maximum leverage
$\lambda_{\mathrm{MAX}}\approx7.5$. It is perhaps a coincidence that $7.5$ is
the maximum leverage allowed for equity trading in the United States, but in
any case this demonstrates that the numbers produced by this model are reasonable.

In Fig.~\ref{price_ts} we show the log-returns $r(t)$ as a function of time.
The case $\lambda_{\mathrm{MAX}} = 1$ is essentially indistinguishable from
the pure noise trader case; there are no large fluctuations and little
temporal structure. The case $\lambda_{\mathrm{MAX}} = 10$, in contrast, shows
large, temporally correlated fluctuations. The autocorrelation function shown
in panel (c) is similar to that observed in real price series. This suggests
that this model may also explain clustered volatility \cite{Engle82}.

\begin{figure}
\begin{center}
\includegraphics[width=11.0cm]{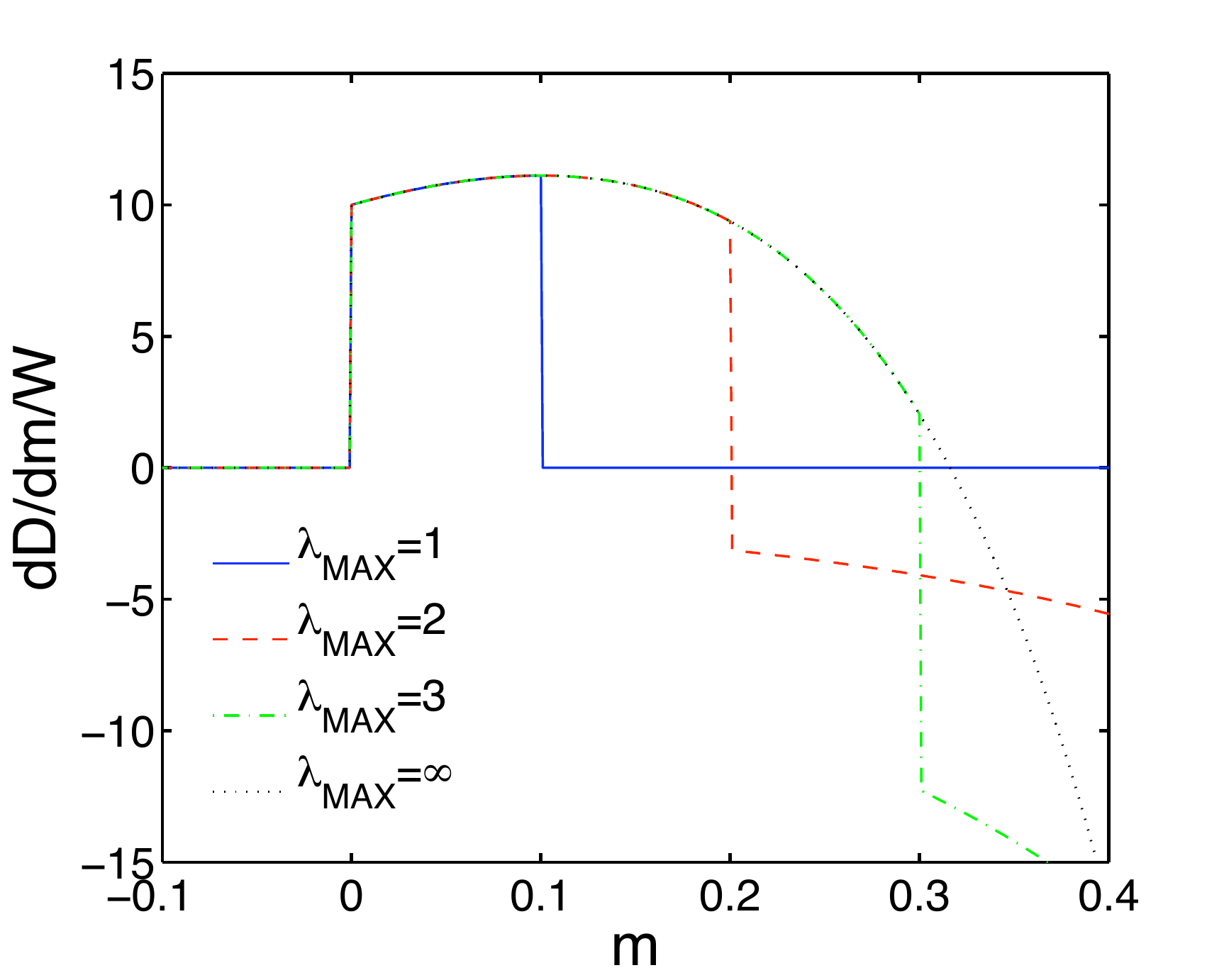}
\end{center}
\caption{Change in demand $dD/dm$ as a result of an infinitesimal increase in the mispricing  from the mispricing $m = V - p$ prevailing last period.  $dD/dm > 0$ means the fund buys when the mispricing increases, and $dD/dm < 0$ means the fund sells when the mispricing increases.  Here the fund has $\beta = 10$ and $W = 1$.  When the maximum leverage $\lambda_{\mathrm{MAX}} =1$ and $m < \lambda_{\mathrm{MAX}}/\beta = 0.1$ the fund responds to an increased mispricing by buying, and when $m > 0.1$ the fund neither buys nor sells (it holds its position).  When $\lambda_{\mathrm{MAX}} =2$ the fund continues to buy until $\lambda_{\mathrm{MAX}}/\beta = 0.2$, but sells for $m > 0.2$.   When $\lambda_{\mathrm{MAX}} =3$ the fund sells heavily for $m > 0.3$.  Finally, when $\lambda_{\mathrm{MAX}} = \infty$,  the fund gradually begins selling anyway; however, as discussed in the text, this latter behavior is a somewhat arbitrary property of the demand function we have chosen.  \label{Demand2}}
\end{figure}

\section{How leverage induces nonlinear feedback}

\subsection{When do the funds sell?}

The fat tails of price movements in our model are explained by the nonlinear
positive feedback caused by leveraging.  When the funds are unleveraged, they will always buy into a falling market, i.e. when the price is dropping they are guaranteed to be buyers, thus damping price movements away from the fundamental value.  When they are sufficiently leveraged, however, this situation is reversed -- they sell into a falling market,  thus amplifying the deviation of price movements away from fundamental value.

This is easily understood by differentiating the fund's demand function given
in Eq. (2-3) with respect to the mispricing. Ignoring the slow moving fund
deposits and redemptions $F(t)$, write $W(t)=D(t-1)p(t)+C(t-1)$. Recalling
that $m=V-p(t)$ and differentiating gives
\begin{eqnarray*}
\mbox{For } m < m^{crit}:~~dD/dm & = & \beta \left[ D(t-1) + \frac{C(t-1) V}{(V-m)^2} \right],\\
\mbox{For } m > m^{crit}:~~dD/dm & = & \frac{\lambda_{\mathrm{MAX}} C(t-1)}{(V-m)^2}.
\end{eqnarray*}
As long as the fund always remains unleveraged, the cash $C(t-1)$ is always positive and the derivative of the demand with the mispricing is always positive.  This means the fund always buys as the price is falling.  In contrast, when the fund is leveraged then $C(t-1)$ is negative.  This means that the fund is always selling as the price is falling when it is above its leverage limit, and depending on the circumstances, it may start selling even before then. 

To visualize this more clearly, differentiate at the point $m=V-p(t-1)$ of
mispricing at the last period. At that point, ignoring redemptions, we can assume
that $D(t-1)$ and $C(t-1)$ are chosen so that for $m<m^{crit}$ the fraction of
the wealth held in the asset is $Dp/W=\beta m$ and the fraction held in cash
is $C/W=1-\beta m$. Similarly, if the fund is over its leverage limit we can
assume that the fraction of the wealth held in the asset is $Dp/W=\lambda
^{\mathrm{MAX}}$, and the fraction held is cash is $C/W=1-\lambda
^{\mathrm{MAX}}$. This implies that the rate of buying or selling under an
infinitesimal change in the mispricing from last period is
\begin{eqnarray*}
\mbox{For } m < m^{crit}:~~ \frac{dD/dm}{W} & = & \frac{\beta( V - \beta m^2)}{(V-m)^2},\\
\mbox{For } m > m^{crit}:~~\frac{dD/dm}{W} & = & \frac{\lambda_{\mathrm{MAX}} (1- \lambda_{\mathrm{MAX}})}{(V-m)^2}.
\end{eqnarray*}
When the fund is leveraged then $1 - \lambda_{\mathrm{MAX}} < 0$, and the second 
term is negative, so when $m > m^{crit}$ the fund always sells  as the mispricing 
increases.  If $\beta m^2 > V$ then the fund may sell as the mispricing increases even when $m < m^{crit}$.

This is illustrated in Fig~\ref{Demand2}, where we plot the derivative of the fund's demand function, $dD/dm$, as a function of the mispricing $m$.  First consider the case where the maximum leverage is one ($\lambda_{\mathrm{MAX}} = 1$).  The fund buys as the mispricing increases as long as the mispricing is small enough that the leverage is under the leverage limit, i.e. for $m < m^{crit} = \lambda_{\mathrm{MAX}}/\beta = 0.1$.  When the mispricing becomes greater than this it simply holds its position. In contrast, with a maximum leverage of two the critical mispricing increases to $m^{crit} = 0.2$.   
The fund now buys as the mispricing increases over a wider range of mispricings, but
switches over to selling when $m>m^{crit}$. When the leverage is further
increased to three, this effect becomes even stronger, i.e. the fund sells
even more aggressively while the price is falling.

Even when there is no cap on leverage, for a sufficiently large mispricing the fund eventually becomes a seller as the mispricing increases.  This is a consequence of the fact that we chose the demand function to be proportional to wealth.  When the mispricing becomes large enough the decrease in wealth overwhelms the increase in the mispricing, so the fund sells even without a margin call from the bank.  This can be viewed as a kind of risk reduction strategy on the part of the fund.  The choice of this demand function is fairly arbitrary, however; we could have implemented an alternative demand function in which the fund never sells when the price is falling, and if it goes out of business, it gives its shares back to the bank without selling them\footnote{This actually happened when the Bear-Stearns hedge funds went out of business; the bank attempted to sell the underlying assets, but the liquidity was so low that they gave up and simply held them.}.  In this case there would be no nonlinear feedback, and thus no systemic risk effect.

\subsection{Nonlinear amplification of volatility}

The key point is that if the fund is leveraged, once the mispricing becomes great enough it transitions from being a buyer to being a seller.  When the fund is below the leverage limit it damps volatility, for the simple reason that it  buys when the price falls, opposing and therefore damping price movements.  It is easy to show that with a reasonably low leverage limit $\lambda_{\mathrm{MAX}}$, when $\lambda < \lambda_{\mathrm{MAX}}$ the expected volatility $E[r_{t}^{2}]$ is damped by a factor approximately
$1/(1+\frac{\beta}{N}(C_{h}+D_{h}V))<1$ relative to the volatility for noise traders alone, where $N$ is the total number of shares of the asset. 

\begin{figure}
\begin{center}
\includegraphics[width=8.0cm]{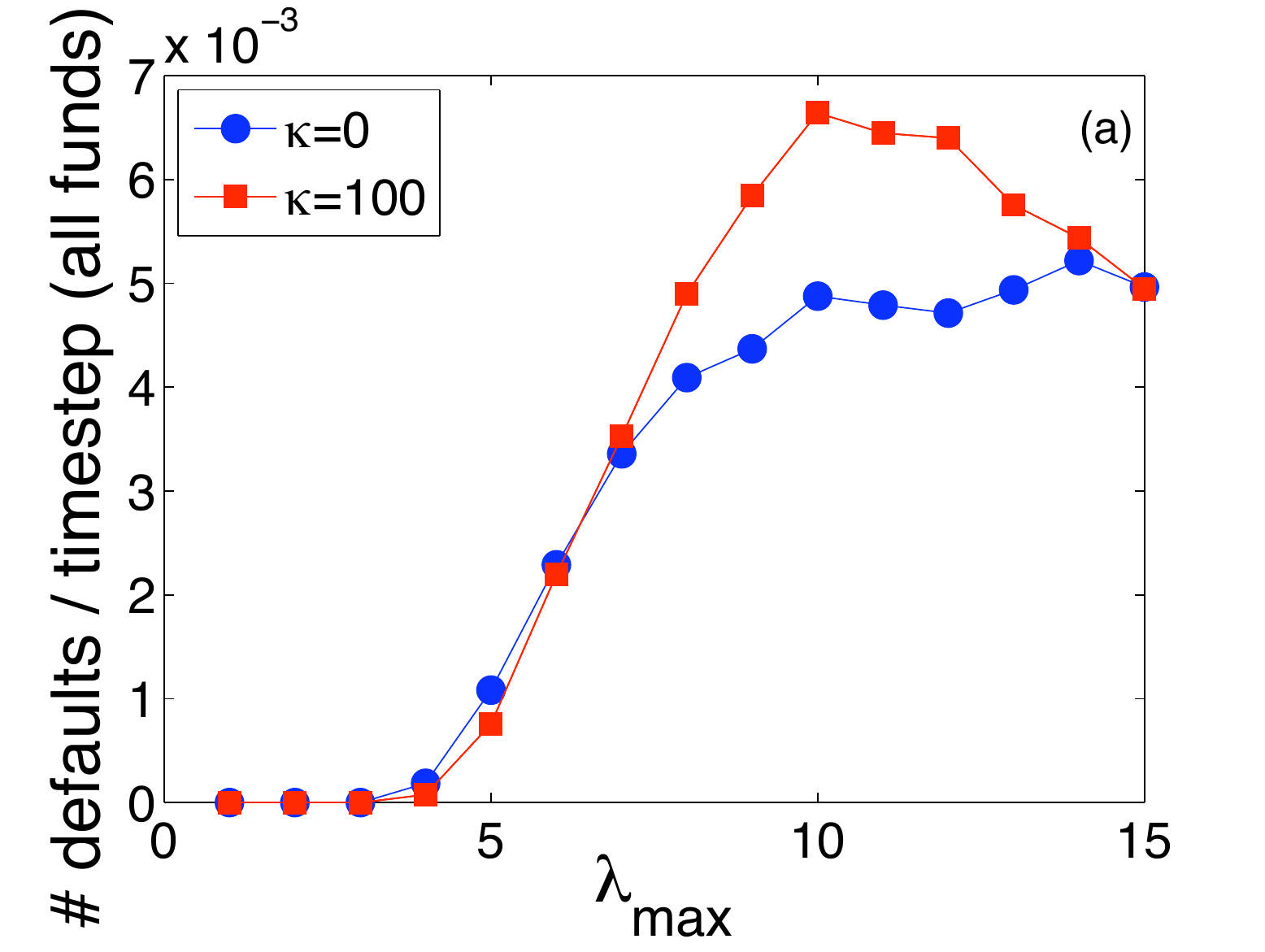}
   \includegraphics[width=8.0cm]{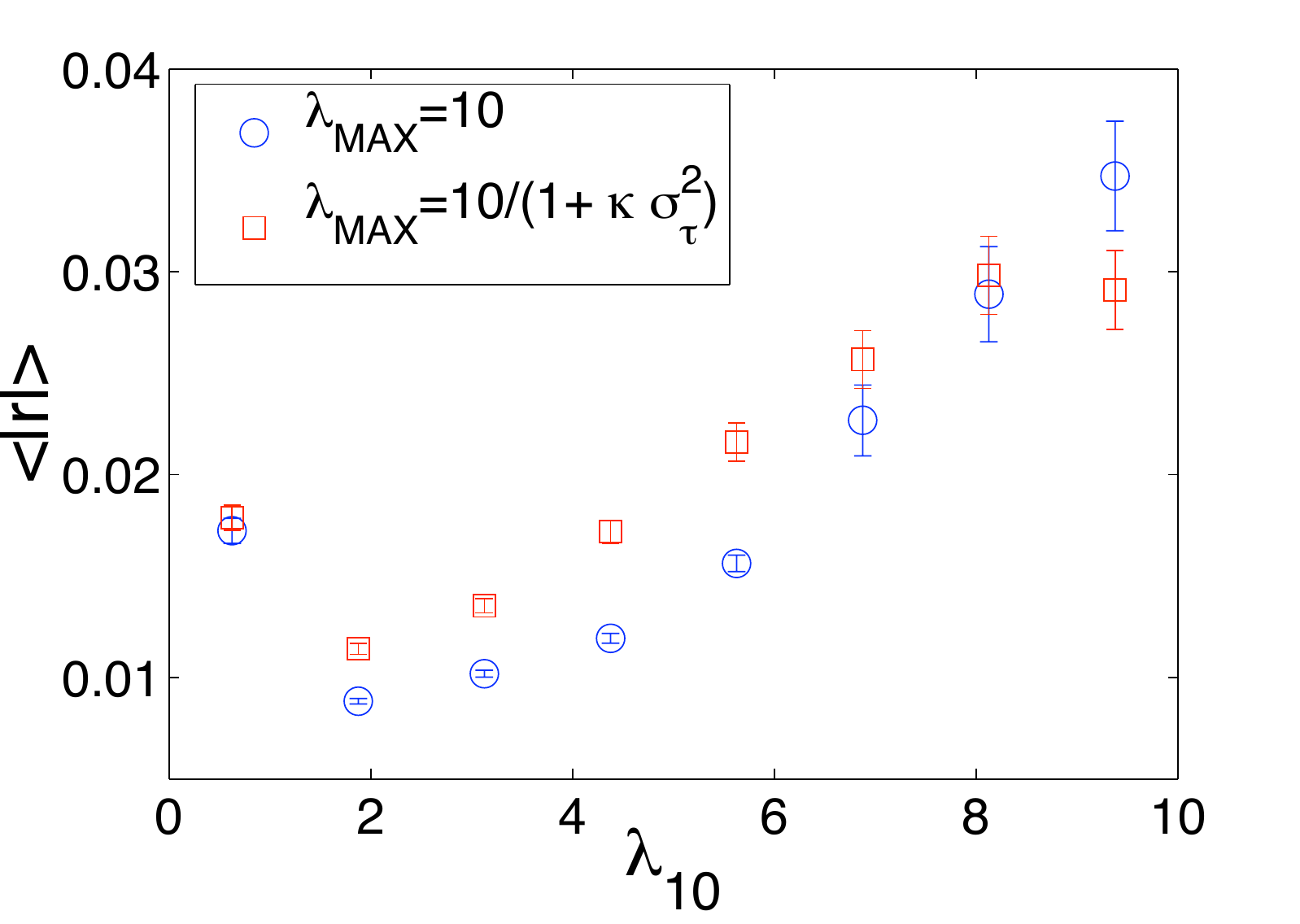} 
  \includegraphics[width=8.0cm]{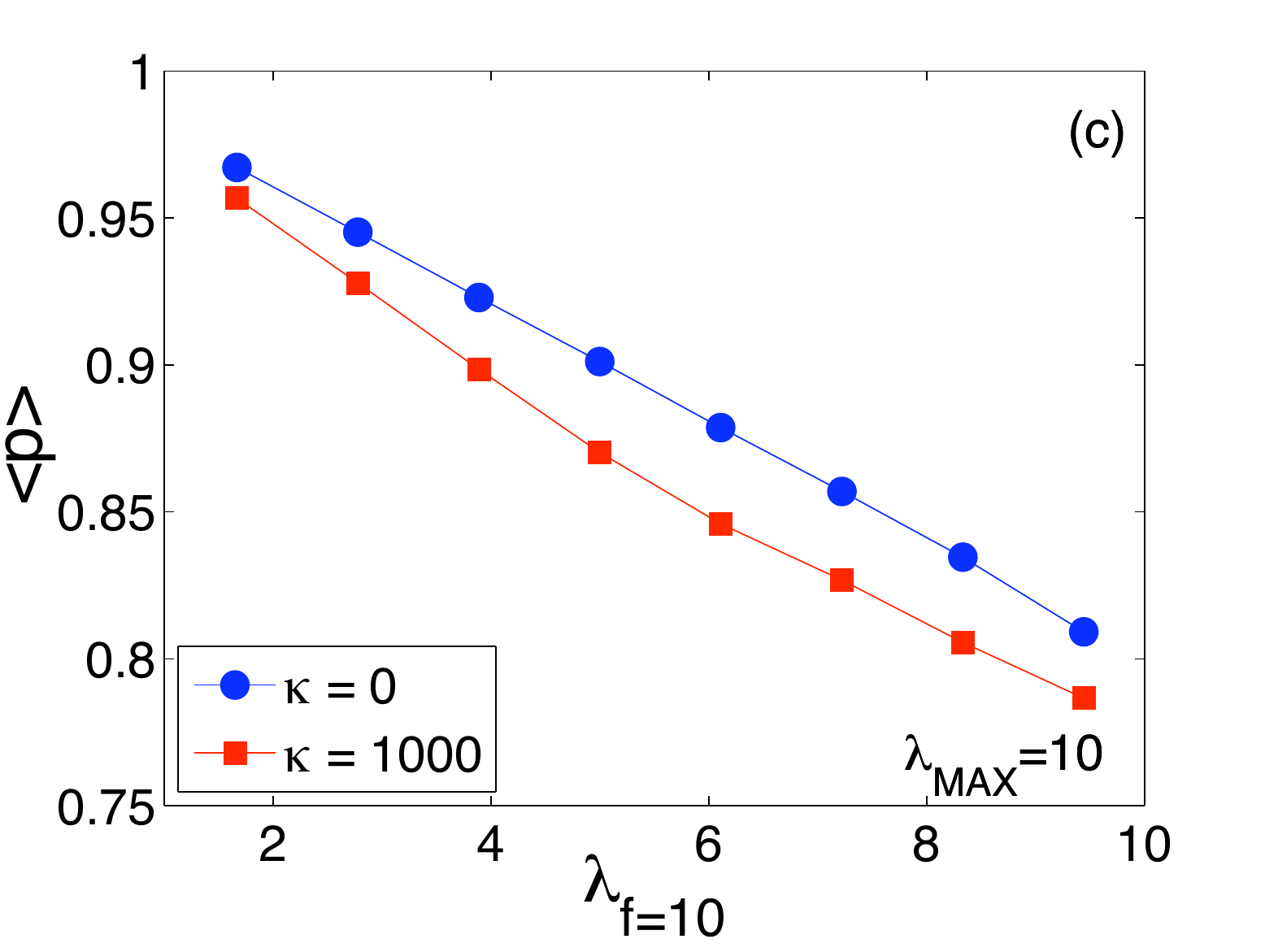} 
\end{center}
\caption{Use of a volatility dependent leverage can increase defaults, increase volatility, and drive prices further away from fundamentals.  The maximum leverage $\lambda_{\mathrm{MAX}}$ is either held constant (blue circles) or varied (red squares) according to 
$\lambda_{\mathrm{adjust}}(t) = \max \left[ 1, \frac{ \lambda_{\mathrm{MAX} } }{1 + \kappa  \sigma^2_\tau }  \right] $, where $\kappa$ is the bank's responsiveness to volatility, and $\sigma^2_\tau$ is the asset price variance computed over an interval of $\tau$ timesteps.  There are 10 funds with the same $\beta$ values as in Figure 2; $\kappa=100$ and $\tau=10$.  Panel (a) shows the default rates as the maximum leverage varies.  In (b) and (c) the maximum leverage is fixed at $\lambda_{\mathrm{MAX}} = 10$.  Panel (b) shows the volatility computed as the average absolute value of logarithmic price returns and (c) the average price.  Note that the variable maximum leverage policy performs more poorly even though the maximum leverage is always less than or equal to its value under the fixed leverage policy.
\label{volRegulation}}
\end{figure}

When funds reach their maximum leverage this reverses and funds instead
amplify volatility. To remain below $\lambda_{\mathrm{MAX}}$ the fund is
forced to sell when the price falls. The volatility in this case is amplified
by a factor approximately $1/(1-\frac{\lambda_{\mathrm{MAX}}}{N}V)>1$. This
creates a positive feedback loop: Dropping prices cause funds to sell, which
causes a further drop in prices, which causes funds to sell. This is clearly
seen in Fig.~\ref{price_ts}(b), where we have placed red triangles whenever at
least one of the funds is at its maximum leverage. All the largest negative
price changes occur when leverage is at its maximum. 
The amplification of volatility by leverage is illustrated in Fig.~\ref{volRegulation}(b) where we show that the average volatility is an increasing function of the average leverage used by the most aggressive fund.  

Thus we see that under normal circumstances where the banks impose leverage limits, the proximate
cause of the extreme price movements is the margin call, which funds can meet
only by selling and driving prices further down. Of course we are not saying
banks should not maintain leverage at a reasonable level; we are only saying
that if they all maintain leverage at a similar level, many funds may make
margin calls at nearly the same time, inducing an instability in prices.  As we have already pointed out, this can be averted by using alternative risk control policies.

\begin{figure*}
 \begin{center}
   \includegraphics[width=13.5cm]{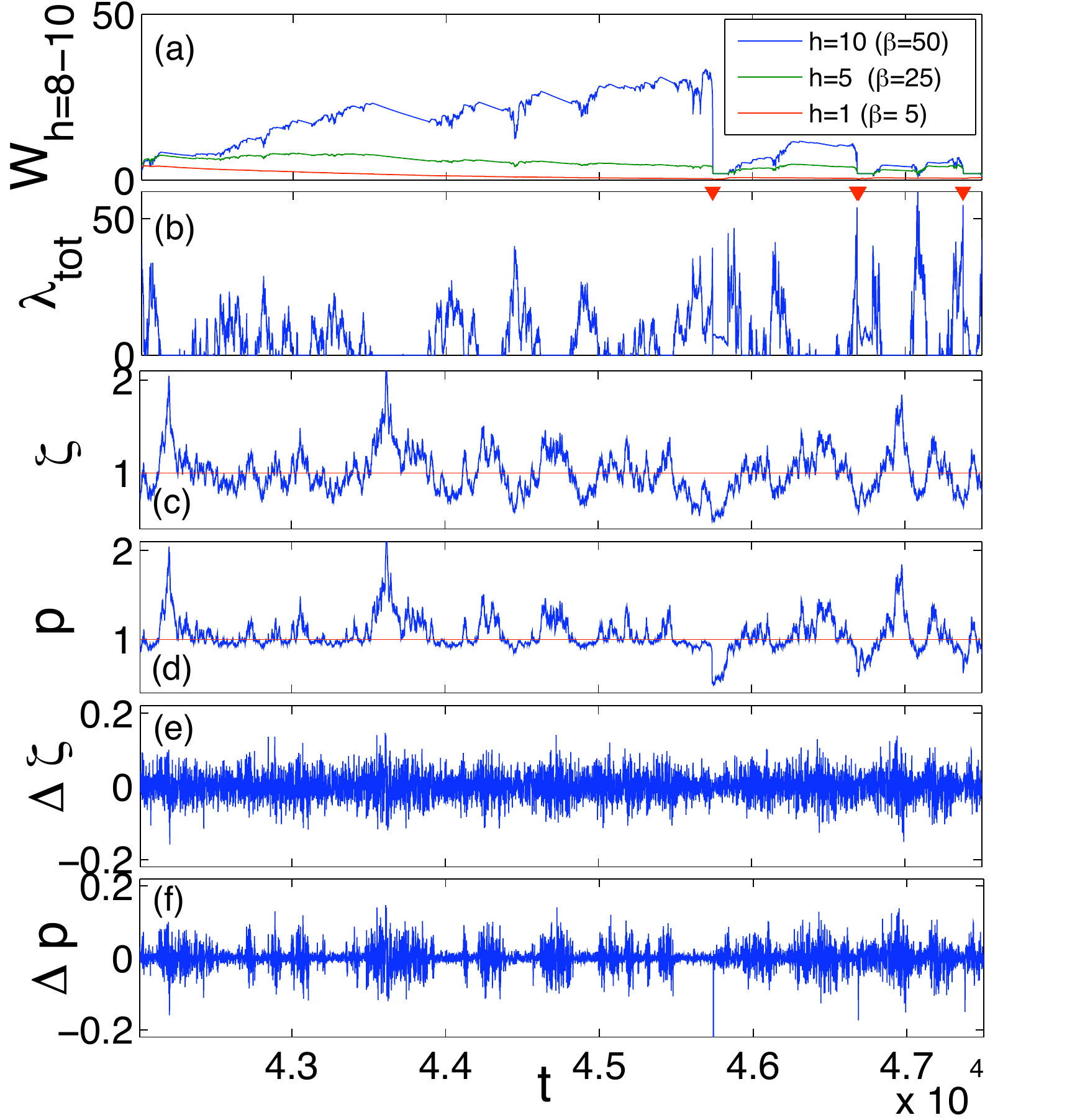}
 \end{center} 
 \caption{Anatomy of a crash.  From top to bottom we see (as a function of time):  (a) The wealth of three representative funds whose aggression levels $\beta_h$ range from highest to lowest. (b) The average leverage, calculated by summing  the demand and wealth in Eq.~\ref{leverageEQ} over all funds.  (c) The noise trader demand $\xi$.  (d) The price.  (e) The fluctuations in the noise trader demand, $\Delta \xi$.  (f) The change in price on successive timesteps, $\Delta p$.  As described in the text, this illustrates how the buildup of leverage causes a crash.  We also see a clear demonstration of how the small clustered volatility in $\xi$ is enormously amplified in the price. }
 \label{storyFig}
\end{figure*} 

\begin{figure*}
 \begin{center}
 \begin{tabular}{cc}
  \includegraphics[width=8.0cm]{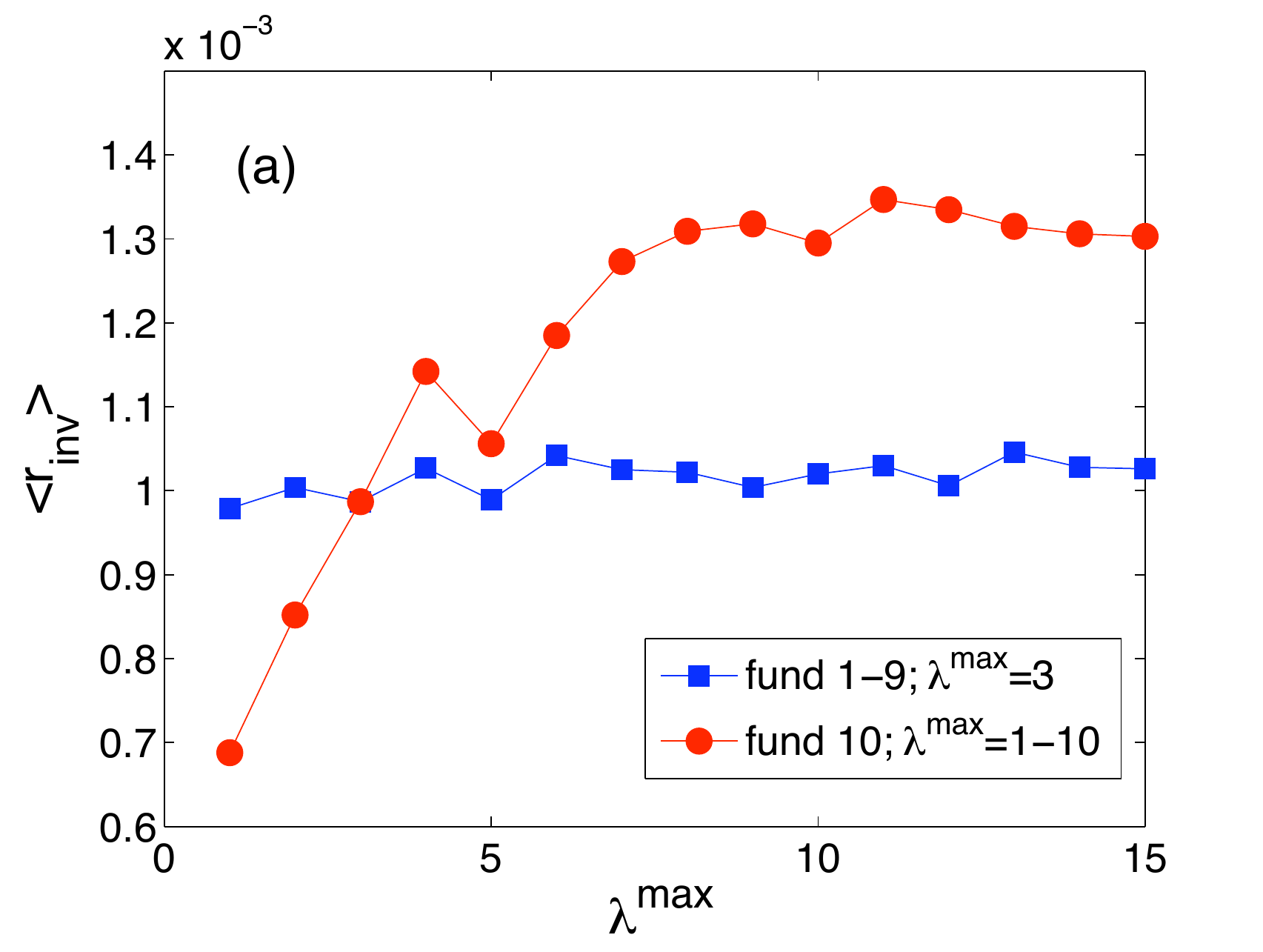} &
 \end{tabular}
 \end{center} 
 \caption{Demonstration of evolutionary pressure to increase leverage.  The maximum leverage is held constant at $\lambda^{\rm MAX} = 3$ for nine funds while $\lambda^{\rm MAX}$ varies from $1-10$ for the remaining fund.   The vertical axis shows the returns to investors, with $\beta=20$ for all funds and $T_{\rm wait}=10$.  The fund with a higher leverage limit gets better returns, and so attracts more capital.  Averages are taken over 50 independent simulations of 100,000 timesteps each.}
 \label{evolutionaryPressure}
\end{figure*} 

\subsection{Evolutionary pressure to increase leverage}

There is also longer term evolutionary pressure driving leverage up which comes from the wealth accumulation process in this model, as illustrated in Figure~\ref{storyFig}.   More aggressive funds use higher leverage.   During times when there are no crashes, more aggressive funds make better returns, attract more investment, and accumulate more wealth, and are thus selected over less aggressive funds.  Thus on average, during good times the average leverage used by the funds tends to increase, until there is a crash, which preferentially wipes out most the aggressive funds and resets the average leverage to a lower level.  The wealth dynamics are illustrated in the top panel of Figure~\ref{storyFig}.  The total leverage is shown in the panel below it.  The leverage comes in bursts as mispricings develop, but the size of these bursts tends to get bigger as the relative wealth of the more aggressive funds increases.  

The next panel illustrates the noise trader demand, which is a weakly mean-reverting random walk, and the panel below it illustrates the price.   During positive excursions of the noise trader demand the asset is overpriced, the funds stay out of the market, and the price is equal to the noise trader demand (measured in dollars).  When the noise trader demand becomes negative the asset is underpriced, and under normal circumstances the fund demand prevents the price from moving very far below $V$.  This is true until the first crash occurs, where the floor of demand provided by the funds collapses due to margin calls, and the funds sell instead of buy as the price falls, temporarily driving the price down even faster than where it would be with noise traders alone.

Crashes are typically not caused by unusually large fluctuations in noise trader demand, as shown in the next two panels.  When the crashes occur the value of change in the noise trader demand, $\Delta \xi(t) = \xi(t) - \xi(t-1)$ is nothing out of the ordinary, and indeed for the examples given here it is not even one of the larger values in the series.  Nonetheless,  the associated change in price, $\Delta p(t) = p(t) - p(t-1)$ is highly negative.  Also one can see that, while there is a very small amount of clustered volatility due to the mean reversion in the demand fluctuations of the noise traders, this is enormously amplified in the price fluctuations.

To illustrate the evolutionary pressure toward higher leverage explicitly, we have done simulations holding all but one fund at a constant leverage $\lambda_{\rm MAX}$ and sweeping the maximum leverage of the last fund, as illustrated in Fig.~\ref{evolutionaryPressure}.  For example, if the nine funds have $\lambda_{\rm MAX} = 3$, a fund with $\lambda_{\rm MAX} > 3$ generates higher returns, as seen in the figure, and thus accumulates wealth and becomes the dominant fund.  In a real world situation this would of course put pressure on other fund managers to increase their leverage.  There is thus evolutionary pressure driving leverage up.

\subsection{Better individual risk control can backfire for the system as a whole}

In an attempt to achieve better risk control, banks often vary the maximum
leverage based on the recent historical volatility of the market, lowering
maximum leverage when volatility has been high and raising it when it has been low. This
is prudent practice when lending to a single fund.   But this can be counterproductive when all the funds might be deleveraged at the same time.  Figure~\ref{volRegulation} shows that if all the lenders refine their policy by tightening their leverage restrictions whenever they see increased volatility over some recent window in the market, they cause more aggregate defaults rather than less, at least when the maximum leverage is between 7 and 15.  The reason for this is simple: Lowering
the maximum leverage across all funds can cause massive selling at just the
wrong time, creating more defaults rather than less. Once again, an attempt to
improve risk control that is sensible if one bank does it for one fund can
backfire and create more risk if every bank does it with every fund.

This kind of policy also has another important unintended consequence.  During times of low volatility leverage goes up.  This in turn drives volatility up, which forces leverage back down.  Thus, in such a situation there are stochastic oscillations between leverage and volatility which on average drive volatility up and drive prices further away from fundamentals.  This is illustrated in Fig.~\ref{volRegulation} (b) and (c), where we plot the average volatility and the average price as a function of the leverage of the most leveraged fund.  Note that the amplification of volatility occurs even though, under the variable maximum leverage risk control protocol, the maximum volatility is always less than or equal to its value under the fixed protocol.

\section{Conclusion}

The use of leverage in the economy is not just an esoteric matter relating to
funds: It is unavoidable. It is the mechanism through which most people are
able to own homes and corporations do business. Credit (and thus leverage) is
built into the fabric of society. The current financial crisis perfectly
illustrate the dangers of too much leverage followed by too little leverage.
Like Goldilocks, we are seeking a level that is \textquotedblleft just right".  This raises the question of what that level is \cite{Peters09}.

We are not the first to recognize the downward spiral of margin calls. After
the Great Depression the Federal Reserve was empowered to regulate margins and
leverage. The model we have developed here provides a quantifiable
framework to explore the consequences of leverage and its regulation. Recent
empirical work has found a correlation betweeen leverage and volatility
\cite{Adrian08}, but our work suggests a more subtle relationship. We make the
falsifiable prediction that high leverage limits, such as we had in reality
until very recently, cause increased clustering of volatility and fat tails,
and that these effects should go up and down as leverage goes up and down. During good times leverage tends to creep up, creating a dangerous situation leading to a sudden crash in prices.  We
have shown that when individual lenders seek to control risk through adjusting
leverage, they may collectively amplify risk. Our model can be used to search
for a better collective solution, perhaps coordinated through government regulation.

At a broader level, this work shows how attempts to regulate risk at a local
level can actually generate risks at a systemic level. The key element that
creates the risk is the nonlinear feedack on prices that is created due to
repaying loans at a bad time. This mechanism is actually quite general, and
also comes into play with other risk control mechanisms, such as stop-loss
orders and many types of derivatives, whenever they generate buying or selling
in the same direction as price movement. We suspect that this is a quite
general phenomenon, that occurs in many types of systems whenever optimization
for risk reduction is done locally without fully taking collective phenomena
into account.


\end{document}